\documentclass[superscriptaddress,aps,pra,longbibliography,dvipsnames,twocolumn,xcolor=table]{revtex4-1}
\usepackage[main=english]{babel}
\usepackage{dcolumn}
\usepackage{bm}
\usepackage{siunitx}
\usepackage{dsfont}
\usepackage{caption}
\usepackage{amsthm}
\usepackage{mathtools}

\usepackage{comment}
\usepackage{multirow}
\usepackage{soul} 
\usepackage{epigraph}

\usepackage{mathtools}
\usepackage{ragged2e}

\usepackage[utf8]{inputenc}
\usepackage[toc,page]{appendix}

\usepackage{graphicx}
\usepackage{hyperref}
\hypersetup{colorlinks}
\usepackage{soul}
\usepackage{bbold}
\usepackage{amssymb}
\usepackage{float}
\usepackage{hyphenat}

\newcommand{\update}[1]{{\color{black} #1}}

\PassOptionsToPackage{dvipsnames}{xcolor}
\RequirePackage{xcolor} 
\definecolor{RoyalBlue}{cmyk}{1, 0.50, 0, 0}
\hypersetup{colorlinks=true,citecolor=ForestGreen,linkcolor=BrickRed,filecolor=blue,urlcolor=RoyalBlue,breaklinks=true}

\begin{document}

\newcommand{\orcidicon}[1]{\href{https://orcid.org/#1}{\includegraphics[height=\fontcharht\font`\B]{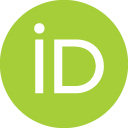}}}

\author{Oriel~Kiss~\orcidicon{0000-0001-7461-3342}}
\email{oriel.kiss@unige.ch}
\affiliation{European Organization for Nuclear Research (CERN), Geneva 1211, Switzerland}
\affiliation{Department of Nuclear and Particle Physics, University of Geneva, Geneva 1211, Switzerland}

\author{Daniil Teplitskiy~\orcidicon{0009-0005-2911-5260}}
\affiliation{European Organization for Nuclear Research (CERN), Geneva 1211, Switzerland}

\author{Michele~Grossi~\orcidicon{0000-0003-1718-1314}}
\affiliation{European Organization for Nuclear Research (CERN), Geneva 1211, Switzerland}

\author{Antonio~Mandarino~\orcidicon{0000-0003-3745-5204}}
\email{antonio.mandarino.work@gmail.com}
\affiliation{International Centre for Theory of Quantum Technologies, University of Gdańsk, Jana Bażyńskiego 1A, 80-309 Gdańsk, Poland}
\affiliation{Department of Physics Aldo Pontremoli,Università degli Studi di Milano, Via Celoria 16, 20133 Milano, Italy}

\title{Statistics of topological defects across a phase transition in a digital superconducting quantum processor}


\date{\today}

\begin{abstract} 
When a quantum phase transition is crossed within a finite time, critical slowing down disrupts adiabatic dynamics, resulting in the formation of topological defects. The average density of these defects scales with the quench rate, adhering to a universal power law as predicted by the Kibble-Zurek mechanism (KZM). In this study, we aim to investigate the counting statistics of kink density in the 1D transverse-field quantum Ising model. We demonstrate \update{ on multiple quantum processing units up to 100 qubits, }that higher-order cumulants follow a universal power law scaling as a function of the quench time. We also show the breakdown of the KZM for short quenches for finite-size systems. Tensor network simulations corroborate our quantum simulation results for bigger systems not in the asymptotic limit. 
\end{abstract}

\maketitle

\section{Introduction}
The Kibble-Zurek mechanism (KZM) is a fundamental theory in nonequilibrium statistical physics, widely used to describe the dynamics of systems undergoing continuous phase transitions. Initially motivated by cosmological considerations regarding structure formation in the early Universe \cite{KIBBLE1980183}, Kibble's pioneering work laid the groundwork for Zurek to extend these ideas to superfluid helium \cite{Zurek1985}, paving the way to cosmological experiments via simulation in acoustic analogues. According to the KZM, when a system is driven across a phase transition, by varying a (or a set of) time-dependent control parameter $h(t)$ which crosses its critical value $h_c$ within a finite quench time $\tau_Q$, it cannot remain in equilibrium due to the critical slowing down phenomenon \cite{dziarmaga2010dynamics, del2014universality}.

At the critical point $h_c$, systems showing a second-order phase transition exhibit a null mass gap or a divergent correlation length, therefore their ground state does not evolve any longer adiabatically \cite{dziarmaga2010dynamics, RMP_Neq}. As the system approaches the critical point, the equilibrium relaxation time diverges, preventing the system from adapting quickly enough to remain in equilibrium. This results in non-adiabatic dynamics, leading to the formation of excitations such as topological defects. These defects, which form at the boundaries between independently developed domains—each selecting different broken symmetries—are robust remnants of the original symmetries.

The KZM predicts a universal scaling law for the mean value of the density of defects, $n$, as a function of the quench time, namely $n \propto \tau_Q^{-\alpha}$. The exponent $\alpha$ is determined by the system's geometry and by the two critical exponents $z$ and $\nu$, the dynamical and correlation length ones respectively, such that the logarithm of the mass gap scales as the product of the two. This scaling behavior has been confirmed in both classical and quantum systems, making the KZM a robust tool for analyzing nonequilibrium dynamics \cite{Lukin_experiment, Campo_20_experiment}. The defect density deviates from the predictions of the KZM in both the fast and slow quench limits. For nearly instantaneous quenches ($\tau_Q < 1$), defects do not have sufficient time to form, resulting in a constant defect density. Conversely, in the slow-driving regime, the dynamics approach the adiabatic limit, where the defect density exhibits an exponential decay, becoming significant when $\langle \hat{n} \rangle < 1/N$.

In the quantum realm, the KZM has been validated in systems undergoing quantum phase transitions, such as the transverse field quantum Ising model in one dimension (TFQIM), for which the exact dynamics of a quantum phase transition was analytically derived and explained in terms of a series of Landau-Zener transitions \cite{Jacek_05}. The peculiarities of such systems are that in contrast to the original cosmological derivation and its superfluid analog, here the underlying symmetry is discrete  ($\mathbf{Z}_2$), and a unitary dissipation-free evolution leads to a non-equilibrium configuration where all cumulants—not just the first one (the mean)—of the probability distribution of the topological defects exhibit a KZM-like scaling. Del Campo and collaborators have shown that in the thermodynamical limit the full counting statistics of kinks in TFQIM reveal universal behavior for slow enough quenches \cite{Campo_2018} and a breakdown of the KZM-scaling with a kink formation dynamics independent on $\tau_Q$, i.e., a constant defect density, for fast quench \cite{Campo_23}. In addition, they studied the influence of symmetry-breaking biases on the scaling \cite{Rams_SB_QPT} and the deviations that occur around equilibrium \cite{Campo_23_deviation}. Although the previous work was conducted analytically on systems with one dimension, \textcite{two_dim_kzm} investigated the KZM in two-dimensional systems using numerical techniques. 

Experiments probing the KZM have been conducted across a variety of platforms, including ultracold gases \cite{KZM_ultracold}, Rydberg atoms \cite{Lukin_experiment}, trapped ions \cite{Campo_20_experiment} and both analog \cite{andersen2024thermalization,king2022coherent} and digital \cite{miessen2024benchmarking} quantum devices based on superconducting qubits. \update{The key difference from previous work is that we use general-purpose digital devices and significantly extend the quench time through noise-agnostic error mitigation protocols.} These experiments often involve measuring the statistical properties of topological defects formed during the phase transition, providing empirical support for the KZM and its broader implications in statistical physics. Ultimately, the KZM remains a cornerstone theory, offering a quantitative framework for understanding the formation of topological defects and the dynamics of symmetry breaking in a wide range of physical systems.

Digital quantum computing is rapidly becoming a promising approach for many-body quantum simulations, offering the ability to efficiently perform highly precise computations on large systems. Although asymptotically optimal methods, based on quantum signal processing \cite{QSP_low17}, hold significant potential, they require substantial advancements in quantum error correction \cite{QEC_Lukin2024,acharya2024QEC} and quantum hardware capabilities. Nonetheless, recent experiments, have successfully demonstrated the practical utility of digital quantum computing \cite{quantum_utility_nature,quantum_utiliy_PRR, farrell2023-schwinger-100qubits,lmg_grossi,kiss2024moments}, using instead quantum error mitigation \cite{Review_QEM} strategies. These experiments bridge the gap towards a fault-tolerant paradigm by pushing the capabilities of current quantum devices.

Here we report an experimental confirmation of the KZM for the cumulants of the kink density distribution via simulation on the 20-qubit digital quantum processing unit based on superconducting transmon qubits IQM Garnet \cite{IQM_garnet}, as well as on the larger 156-qubit devices ibm$\_$aachen and ibm$\_$kingston.
\update{For many years, quantum simulation was only possible using analog platforms. In this paper, we emphasize that simulations on multipurpose superconducting-qubit digital devices, beyond the annealing regime, are now a practical and compelling reality.
At the current stage, we think they offer a more flexible architecture than analog simulators, capable of implementing specific classes of models. 
Despite much research conducted over the last decades, we are still far from an unbiased evaluation of which approach will be the leading one in the future.  
Several challenges are still unsolved, ranging from the fundamental problems posed by the Hamiltonian simulation to the control and characterization of decoherence effects induced by noise. 
Analog and digital quantum simulations each offer distinct advantages and limitations when applied to studying quenched critical systems. Analog simulators, usually implemented in cold atoms or trapped ions, by their nature, can mimic the dynamics of phase transitions, allowing for almost real-time observation of defect formation,  achieving relatively large system sizes. However, they often require precise engineering and control, making it impossible to reach wide ranges in the space of parameters. Digital quantum simulations provide high programmability and fine-grained control over system dynamics, once the unitary evolution is suitably decomposed,  enabling the investigation of diverse models and measurement of specific observables. A practical advantage is available when the dynamical evolution is factorizable directly on a set of native gates of the selected hardware, resulting in shallow circuit depths.
In the near future, a combination of both may be necessary to study the complexities of KZM in non-integrable systems.
To sum up, whereas analog methods are actually suited for exploring large-scale dynamical behavior, digital approaches offer greater flexibility and precision. For such reasons, we think that the strategies are mutually complementary, and it is nowadays necessary to delve into the capabilities offered by the latter platforms. 
}

\section{Methods}

\subsection{The model} 

The one-dimensional transverse-field quantum Ising model is a paradigmatic system in the study of quantum phase transitions and critical phenomena. It consists of a linear chain of spins, each governed by two competing interactions: a transverse field and nearest-neighbor coupling. We consider the TFQIM of the form
\begin{equation}
\label{eq:Ising}
    H(t) = -J(t)\sum^{N-1}_{i=1} X_iX_{i+1} - h(t)\sum^N_{i=1} Z_i 
\end{equation}
with $J(t) = J_0t/\tau_Q$ and $h(t) = (1-t/\tau_Q)h_0$. The competition between these interactions drives a quantum phase transition at a critical field strength $h_c/J_c=1$, where the system undergoes a transition from a ferromagnetic phase, characterized by ordered spin alignment, to a paramagnetic phase, where the spins are aligned with the transverse field. The model is exactly solvable, making it a cornerstone for understanding quantum criticality and non-equilibrium dynamics in low-dimensional systems. We fix $h_0=J_0=1$ such that the quantum phase transition is crossed at the middle of the quench, i.e. at $t=\tau_Q/2$. 
The initial state is chosen as the ground state at $t=0$, and takes the form of the \update{paramagnetic state} with all the spins up $|0\rangle ^{\otimes N} \equiv |\bar{0}\rangle$. To measure the number of topological defects we consider the kink operator

\begin{equation}
\label{eq:n}
\begin{split}
    \hat{n} &= \frac{1}{2N} \sum^{N-1}_{i=1} \left(\mathbb{1} - X_i X_{i+1}\right).
\end{split}
\end{equation}
We note here that our choice of open boundary conditions allows us to compute the two-point operators concurring in the evaluation of the defect density on the entire length of the simulated Ising chain. Periodic boundary conditions with small-sized systems would have resulted in less interesting analysis. 
The cumulants are the coefficients of the series expansion of the moment-generating function of the kink distributions. In the quantum simulation, we focus on the first three, namely: 
\begin{align}
\label{eq:first_cumulants}
\kappa_1 &= \langle \hat{n} \rangle, \, \, 
\kappa_2 = \langle (\hat{n}- \langle \hat{n} \rangle)^2 \rangle =    \langle \hat{n}^2 \rangle - \kappa_1^2\\
\kappa_3 &= \langle (\hat{n}- \langle \hat{n} \rangle)^3 \rangle =    \langle \hat{n}^3 \rangle - 3\kappa_1 \kappa_2 - \kappa_1^3
\end{align}
This operator can be estimated up to error $\epsilon$ by measuring the wavefunction in the Hadamard basis at least $\text{Var}[\hat{n}^k]/\epsilon^2 \leq 1/\epsilon^2$ times. All the correlators can be measured in this basis and can therefore be estimated simultaneously. The main challenge arises for long quench times, since the expectation values are expected to be small, and thus require a higher number of measurements to be detected from statistical noise. However, there is no fundamental problem in estimating expectation values in this way. 

\subsection{Time-dependent Hamiltonian simulation}
The quench is performed by evolving the initial state under a time-dependent Hamiltonian $H(t)$. The time evolution is obtained via the time-ordered unitary evolution stemming from the Hamiltonian as

\begin{equation}
\begin{split}
\mathcal{U}(t_f,t_0) = \mathcal{T} \exp \left(-i \int_{t_0}^{t_f} H(t) dt  \right),
\end{split}
\end{equation}

where $\mathcal{T}$ is the time ordering operator to account for the non-commutativity at different time $[H(t),H(t')] \neq 0$. There exist different ways of implementing $\mathcal{U}(t_f,t_0)$ on digital quantum computers, the most straightforward being the quasistatic approximation \cite{Born1928}. In this framework, the simulation time is sliced into $r$ Trotter steps of size $\Delta t=(t_f-t_0)/r$, in which $H(t) \approx H(k'\Delta t)$, for $ k' = k+1/2$ and $t \in [k \Delta t, (k+1) \Delta t]$. The time-evolution operator can thus be approximated by 
\begin{equation}
\label{eq:static}
    \mathcal{U}(t_f,t_0) = \prod_{k=1}^r \exp \left[ -i\Delta t H(t_0 + k'\Delta t) \right].
\end{equation}
The matrix exponential $\exp(-i\Delta t H)$ is not directly implementable on a digital quantum processor. Therefore, we decompose it into fundamental quantum gates \cite{Miessen2023}, e.g., using the Trotter-Suzuki product-formula (PF) \cite{SUZUKI1990319}, whose error scales polynomial in time, and pre-factor depending on the commutators \cite{PhysRevX_high_trotter}. For simplicity, let us assume that the Hamiltonian breaks down into two groups of pairwise commutating terms, 

\begin{equation}
\label{eq:td-ham}
H(t) = f(t) A + g(t)  B,
\end{equation}
such that $\exp(i A)$ and $\exp(i B)$ are exactly implementable. We use a second-order \update{Trotter expansion}, which is given in the static case by 
\begin{equation} 
\label{eq:suzuki}
\begin{split}
&\exp(-iH t) = \\
\exp(-if(t) At/2) &\exp(-ig(t)Bt) \exp(-if(t)At/2) \\&\update{+\mathcal{O}(t^3)}.
\end{split}
\end{equation}
Merging both approximations \eqref{eq:static} and \eqref{eq:suzuki}, we obtain 
\begin{equation}
\begin{split}
\mathcal{U}(t) &\equiv \mathcal{U}(t_f=t,t_0=0) \\
&=\prod_{k=1}^r \exp[-i f(k'\Delta t) A \Delta t/2] \\
& \cdot \exp[-i g(k'\Delta t) B\Delta t]\exp[-i f(k'\Delta )t A \Delta t/2].
\end{split}
\end{equation}

\begin{figure}
    \centering
    \includegraphics[width=0.8\linewidth]{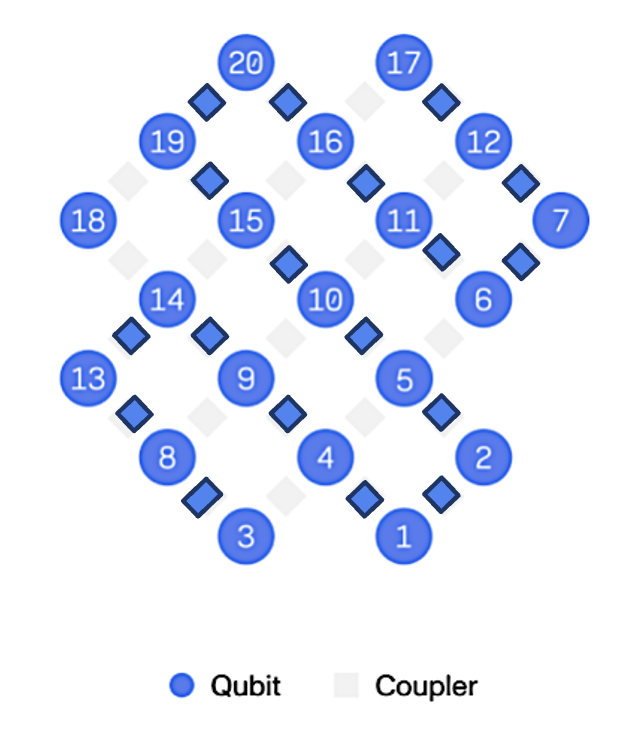}
    \caption{\justifying{\textbf{Topology of IQM Garnet.} The active (inactive) connections used in the experiment are displayed with the shaded blue (gray) squares, and the 18th qubit is inactive since we can not find a full 20-qubit line.}}
    \label{fig:topology}
\end{figure}

Straightforward extensions of the quasistatic PF method include adaptive step sizing \cite{Zhao_Adapt_trotter}, random compilation techniques like QDrift \cite{QDrift,qDRIFT_Kiss} with one-norm scaling \cite{Berry_qdrift_timedependant}, and leveraging the interaction picture \cite{Rajput2022hybridizedmethods} or perturbation theory \cite{THRIFT}. While asymptotically superior methods exist—such as those based on the Dyson series \cite{kieferova2019dyson}, Magnus expansions \cite{sharma2024magnus,casares2024magnus}, discrete clocks \cite{watkins2022time}, or flow equations \cite{Boyan_beyond_static}—the PF method is notably effective in practice, especially for systems with a high degree of locality \cite{PNAS_spins}.

\subsection{Error mitigation}
\label{sec:qem}

\paragraph{Noise renormalization}
The workhorse of our error mitigation protocol is based on a noise renormalization technique, successfully used in \cite{noise_estimation,self-mitigation,farrell2023-schwinger-100qubits,kiss2024moments,Kavaki2024}. The idea is to assume a noise model, estimate the corresponding parameters on the quantum device and reverse its effect in the post-processing stage. 
In the following, we shall assume a depolarising noise model $\mathcal{N}_p[\cdot]$ with parameter $p$, which maps the expectation value of a Pauli observable $\sigma$ under a quantum state $\rho$ to
\begin{equation}
    \text{Tr}\left(\sigma \mathcal{N}_p[\rho]\right)= (1-p)  \text{Tr}\left(\sigma \rho\right).
\end{equation}
If the parameter $p$ would be known, the results can be corrected by dividing the noisy results by $(1-p)$ 
\update{
\begin{equation}
     \text{Tr}\left(\sigma \rho\right)_{\text{corrected}} = \frac{1}{(1-p)} \text{Tr}\left(\sigma \rho\right)_{\text{noisy}}.
\end{equation} }

The main idea is thus to estimate $(1-p)$ by running a circuit with a known expectation value. This becomes particularly simple when $h=0$ or $th(t)=\pi$. \update{Hence, in both cases, the action of the field commutes with the interaction part, i.e. $[\mathcal{U}(t),\hat{n} ]=0$ for $t=0$ or $th(t)=\pi$. This leads to
\begin{equation}
\label{eq-circ-odr}
    \langle \bar{+}|\tilde{\mathcal{U}}^\dagger(t)\hat{n}\tilde{\mathcal{U}}(t) |\bar{+} \rangle  = (1-p),
\end{equation}
where $|\bar{+}\rangle$ is the uniform superposition state obtained by applying a layer of Hadamard gates to $|0\rangle^{\otimes N}$, and $\tilde{\mathcal{U}}(t)$ is the time evolution when setting $th(t)=\pi$. Since we want the circuit to be as close as possible to the original one, we choose the field such that $h(t)=\pi/t$. Therefore, the only difference with the original circuits is the rotation angles of the field, enabling us to effectively estimate $(1-p)$ by evaluating \eqref{eq-circ-odr}.}

\paragraph{Randomized compiling}
The main assumption of the noise renormalization protocol is that the noise is depolarising. However, this is not the case on real quantum devices. To alleviate this effect, the circuits are randomly compiled \cite{randomized_compiling_PRX}. In practice, each two-qubit gate is twirled using single\hyp qubit Pauli rotation, chosen in such a way that the transformed and original circuits are equivalent. This process is know as Pauli twirling \cite{Pauli_Twirling} and effectively turns the noise channel into a Pauli channel. Moreover, under twirling, the noise accumulates as in a random walk \cite{PhysRevA_random_wallman}, which is quadratically slower than it would add coherently. Since the data obtained in this process are merged together before computing expectation values, we can simply distribute the shot budget over all the twirls, thus avoiding an increase in the number of samples. 

Twirling can also be used to mitigate measurement errors \cite{TREX} in the computation of expectation values. This process is performed by randomly flipping the state of the qubit before measurement, and reversing the value of the measured bit, if needed. Effectively, this process approximately diagonalize the readout transfer-matrix, making it easier to invert using the following technique, described below.  

\paragraph{Readout error mitigation}

We mitigate readout errors by calibrating the device. Since a full calibration is exponentially expensive, we adopt a sparse strategy \cite{MEM,Kiss_Li6,efficient_MMM} by only measuring the $N$-qubit state $|0\rangle^{\otimes N}$ and $|1\rangle^{\otimes N}$, and build the confusion matrices
\begin{equation}
P_k=
\begin{pmatrix}
P^{(k)}_{0,0} &P^{(k)}_{0,1} \\P^{(k)}_{1,0} &P^{(k)}_{1,1}
\end{pmatrix},
\end{equation}
where, $P^{(k)}_{i,j}$ is the probability of the $k$-th qubit to be in  $|j\rangle$  while measured in $|i\rangle$, for $i,j\in\{0,1\}$.
The measurements $\vec{M}^k$ of the qubit $k$ is later corrected as 
\begin{equation}
    \vec{M}^k_{\text{corrected}} = (P_k)^{-1}\vec{M}^k.
\end{equation}
As stated above, measurement twirling is performed to approximately diagonalize the transfer-matrix, thus making it easier to invert. 

\begin{figure*}
    
 \centering
    \includegraphics[width=0.95\linewidth]{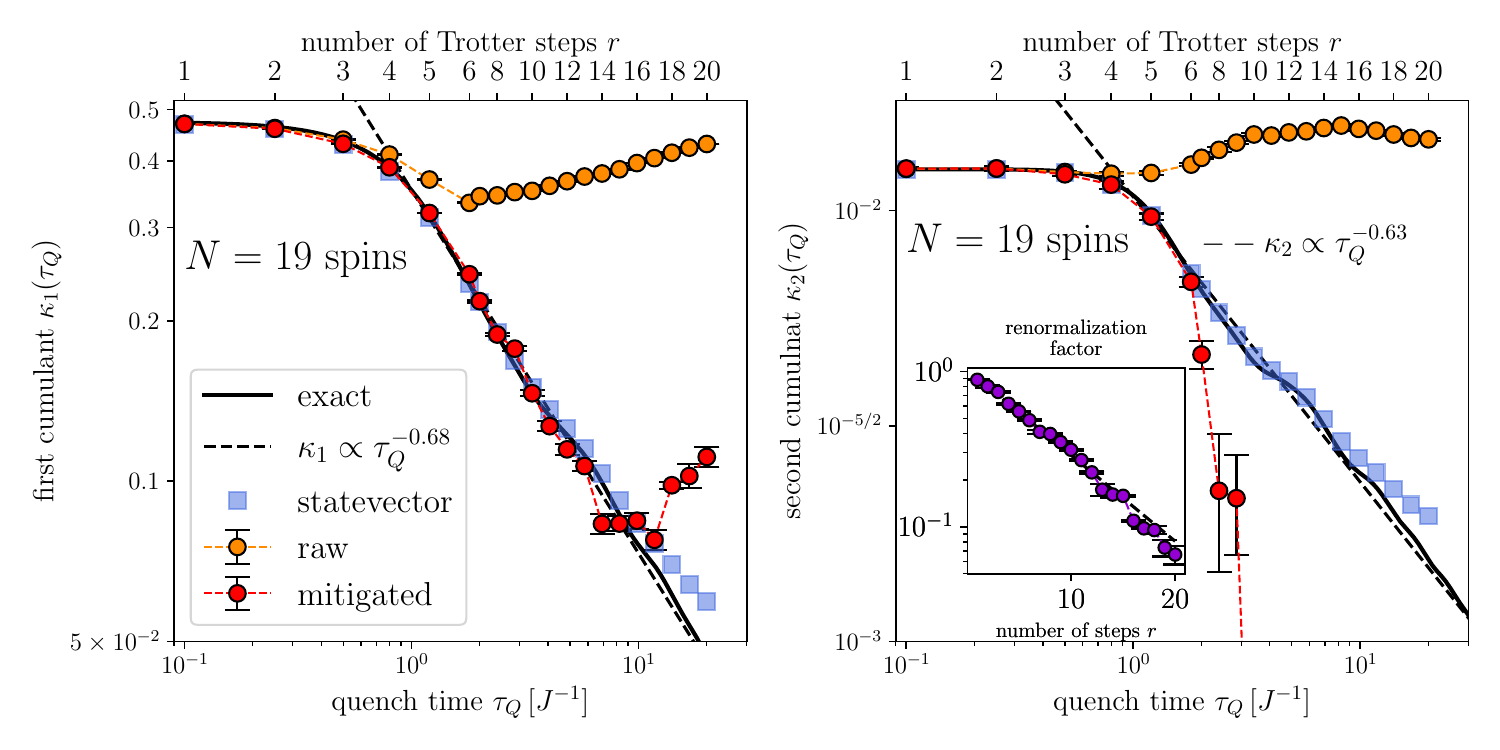}
   
    \caption{\justifying{\textbf{Quantum simulation on IQM Garnet ($N=19$).} The first two cumulants of the kink density are shown as a function of the quench time, as well as the number of Trotter steps. The blue squares denote statevector simulation, the orange points the raw data, and the red the results obtained after the full pipeline of error mitigation. The complexity of the circuits is increased by one additional Trotter step per data point. The black continuous line shows the exact evolution, while the decay rate appears as a dotted line. The inset shows the renormalization factor (purple) as a function of the circuit's depth, resulting from the expression in \eqref{eq-circ-odr}.}}
    \label{fig:harware19}
    
\end{figure*}

\section{Quantum simulation}
In this section, we report numerical experiments, both with quantum hardware \update{(IQM and IBM), statevector simulation (\texttt{qsim} \cite{quantum_ai_team_and_collaborators_2020_4023103}) and tensor networks (\texttt{itensor}),} of the correlation's decay across the quantum phase transition.

\subsection{Experiments on quantum hardware}

We begin by performing the quench on the IQM Garnet superconducting quantum computer \cite{IQM_garnet}, which consists of twenty transmon qubits and achieves a two-qubit gate fidelity of 99.5\% \cite{IQM-PRXQ}. Due to the chip's connectivity, displayed in Fig.~\ref{fig:topology}, we work with a system of $N=19$ spins. We implement the second-order PF method, varying the number of Trotter steps $r$ up to 20 across different quench times $\tau_Q \in [0.1,20]$. The first two cumulants are presented in Fig.~\ref{fig:harware19}, with statevector simulations shown as blue squares, raw data as purple dots, and error-mitigated results as red dots. The continuous line represents a near-exact evolution, obtained on a statevector simulator using 2000 Trotter steps. \update{While this is not an exact result, we choose the size of the Trotter steps such that the error is at most of the order $10^{-6}$.} Higher-order moments are not displayed, as they are several orders of magnitude smaller and would require significantly more measurements to be accurately estimated.

\begin{figure*}
    \centering
    \includegraphics[width=0.95\linewidth]{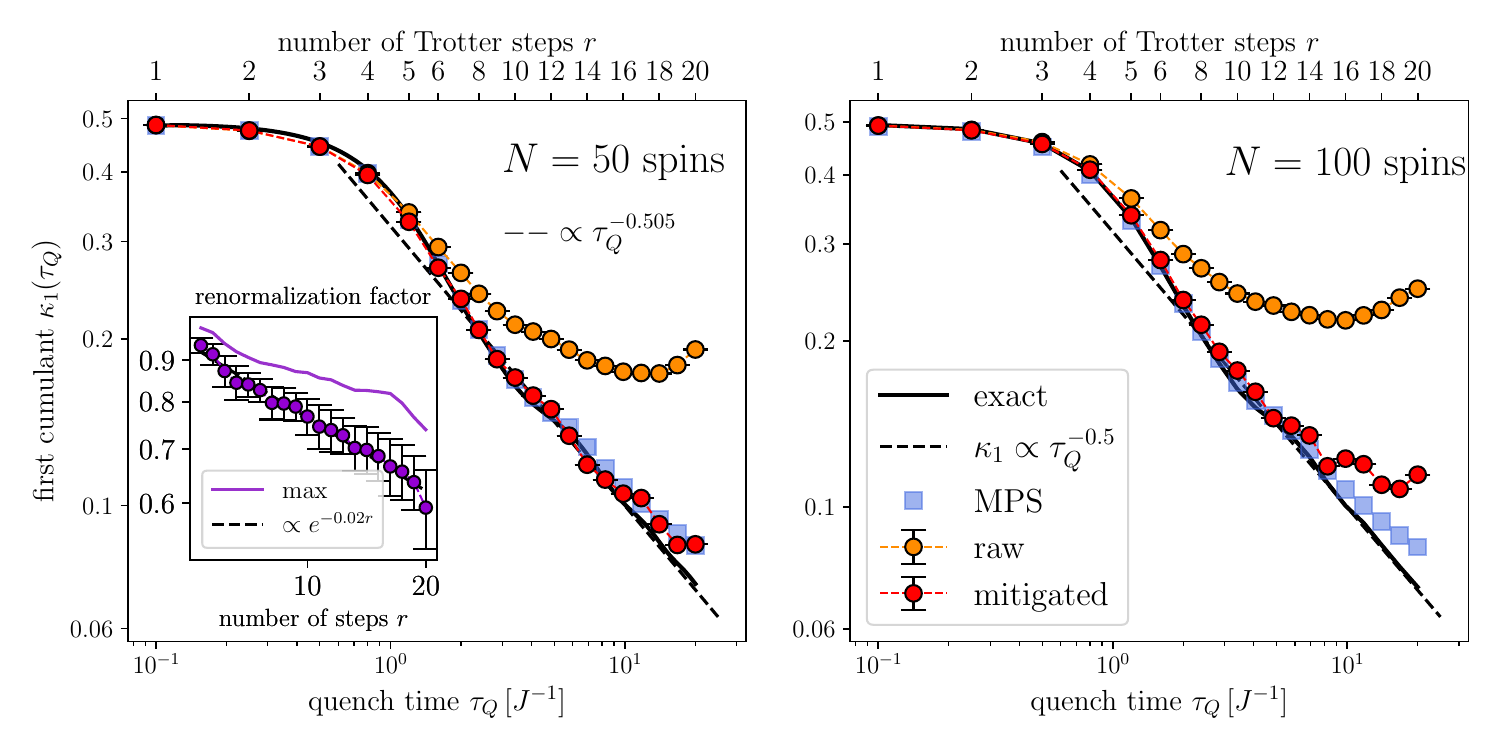}
    \caption{\justifying{\textbf{Quantum simulation on IBM Kingston ($N=50$) and IBM Aachen ($N=100)$.} The first cumulant of the kink density is shown as a function of the quench time, as well as the number of Trotter steps. The blue squares denote MPS simulation of the exact same circuits, the orange points the raw data, and the red the results obtained after the full pipeline of error mitigation. The complexity of the circuits is increased as one additional Trotter step per data point. The black continuous line shows a proxy for the exact evolution, which is obtained with a maximal of 100 Trotter steps, while the decay rate appears as a dotted line. The inset in the $N=50$ plot refers to the renormalization factor computed according to the expression given in (\ref{renormIBM}).}}
    \label{fig:hardware50}
\end{figure*}

\update{We perform a second 50 spins and 100 spins experiment on the ibm$\_$kingston and ibm$\_$aachen Heron r2 devices. They are heavy-hexagonal lattice chips consisting of 156 qubits, with two-qubit gates fidelity between $[99.83\%,99.94\%]$. The results for the first cumulant as shown in Fig.~\ref{fig:hardware50}, and are compared to equivalent MPS simulations. The qubits are selected such as to minimize the worst two-qubit gate fidelity. The second order cumulant can not be resolved because of the limited amounts of samples collected, which should grows as $\mathcal{O}(N^4)$.}

Error mitigation is applied using noise renormalization techniques \cite{noise_estimation,self-mitigation,farrell2023-schwinger-100qubits,kiss2024moments,Kavaki2024}, along with 50 instances each of Pauli \cite{Pauli_Twirling} and readout \cite{TREX} twirling, and matrix-free measurement mitigation \cite{MEM}, as described in the previous Section \ref{sec:qem}. We observe that the renormalization factor decays exponentially fast with the depth, as shown in the inset of the lower plot of Fig.~\ref{fig:harware19}. Therefore, this process can not be applied for an arbitrary-depth circuit, but in this case it enables us to extend it by a factor of four. Each circuits is run 2000 times, and the error bars correspond to a $95\%$ confidence interval computed via Bayesian data augmentation, see Appendix \ref{bayesian} for details. The accuracy limit resulting from finite statistics is $10^{-5/2}$, is depicted on the corresponding axis. 
\update{For the larger experiments, we noted that it was more effective to choose the renormalization factor as the maximum over every observable 
\begin{equation}
\label{renormIBM}
    (1-p) = \max_{0\leq i<N-1} \langle \bar{+}|\tilde{\mathcal{U}}^\dagger(t) X_iX_{i+1}\tilde{\mathcal{U}}(t)|\bar{+}\rangle,
\end{equation}
}
which appears as a solid line in the corresponding inset.

As predicted by the theory, we observe a constant defect density for nearly instantaneous quenches. However, the crossover to the adiabatic regime, expected at slower quenches, is absent. This is attributed to the limitations of current quantum hardware, which lacks the capability to achieve the circuit depth necessary to suppress Trotter errors in this regime. We fit the defect density decay rate for $\tau_Q = 1$ to $\tau_Q = 10$, as the calculations are reliable only within this range, and obtain a decay rate of $0.68$. This value deviates from theoretical predictions due to finite-size effects. 
Due to the concurrent effect of boundary effect and finite numbers of interacting particles, in Fig.\ref{fig:decay_rate} we are able to identify three 
typical regimes in the defect density formation. 
For long quenching times, the correlation length becomes comparable to the system size, 
therefore inducing the system to evolve adiabatically. In this regime, the density of kinks remains nearly zero and is independent of the quenching time. 
On the other extreme, for very short quenching times, the evolution is fully non-adiabatic, causing the density of kinks to reach a maximum value determined by the final quench point. 
Between these two extremes lies the KZ scaling regime, where the density of kinks follows a power-law dependence on the quenching time.
It is worth noting that by the time this paper was submitted an interesting theoretical analysis about the effects related to the role of boundary conditions, endpoints, and eventually system size has been reported in \cite{garcia2024quantumkibblezurekmechanismrole}, that it is in agreement with our findings.

\subsection{Computation with MPS}
To support our findings on finite-size systems, we conduct numerical experiments using matrix-product states (MPS) \cite{MPS-Cirac-2007} with larger spin systems to approximate results obtained in the thermodynamic limit \cite{Campo_2018}. 

We begin by calculating the decay rate $\alpha$ as a function of system size by fitting the data to a model of the form  $\propto \tau_Q^{-\alpha}$, in line with the decay predicted by the KZM. \update{The data used for the fit lies within the range $\tau_Q \in [1, \tau_f]$, where $\tau_f$ is defined such that $\langle \hat{n}(\tau_f) \rangle = 1/N$.}  For this analysis, we perform MPS simulations with a fixed error tolerance of $10^{-10}$, based on which the smallest coefficients (from the singular value decomposition) with a lower sum of their squares than this are discarded, and 300 Trotter steps, extracting the decay rate using $10^5$ shots per run. As previously mentioned, expectation values tend to be small for higher-order cumulants. Therefore, our focus lies on the first three cumulants.
\begin{figure}
    \centering
    \includegraphics[width=0.9\linewidth]{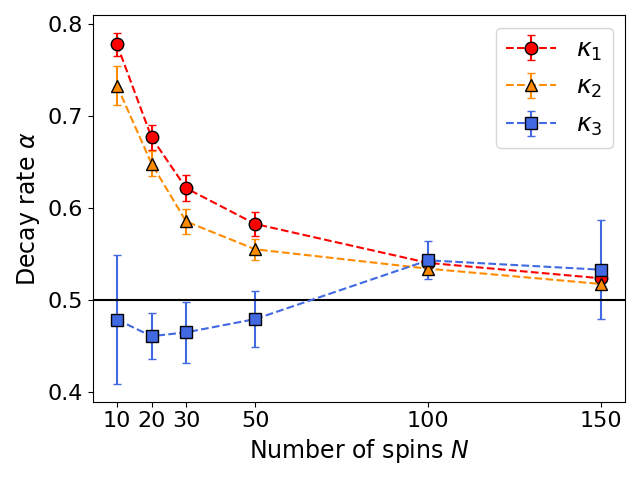}
    \caption{\justifying\textbf{Decay rate scaling.} The decay rate $\alpha$ of the three first cumulants (red, orange and blue respectively) extracted by least-square regression is shown as a function of the system size. The simulations are performed using MPS at fixed error threshold of $10^{-10}$. The error bars correspond to one standard deviation due to the finite statistics. We observe that all decay rates approaches $1/2$ (continuous black line), as predicted by the theory in the thermodynamic limit.}
    \label{fig:decay_rate}
\end{figure}
We show in Fig.~\ref{fig:decay_rate} that the decay rate for the first three cumulants approaches $1/2$, as predicted by \textcite{Campo_2018}. The uncertainty on higher cumulants increases, due to statistical noise. 

\begin{figure}
    \centering
    \includegraphics[width=0.9\linewidth]{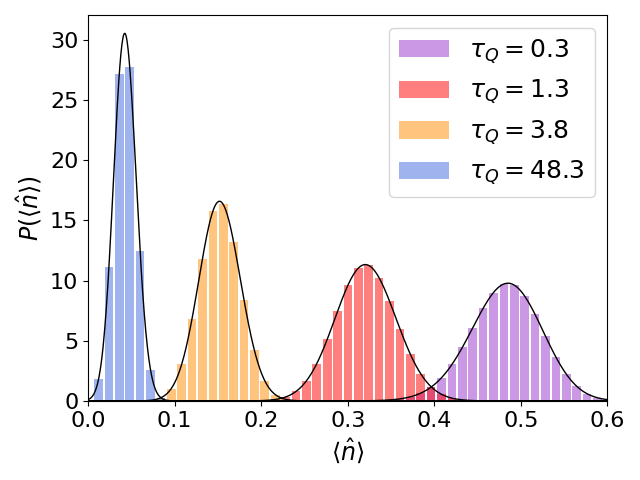}
    \caption{\justifying\textbf{Distribution of the kink density.} The probability distribution function $P(\langle \hat{n}\rangle)$ of the kink density is shown as a function of different quench times (colors). The simulation is performed using MPS on $N=150$ spins.}
    \label{fig:pdf}
\end{figure}

Finally, we reconstruct the kink probability distribution $P(\langle \hat{n}\rangle)$ from the first three moments at fixed system size $N=150$ for different quench times by using the maximum entropy method \cite{maxent}, which is shown in Fig.~\ref{fig:pdf}.
We observe that the probability distributions become narrower, and the expectation value of the kink density decreases with longer quench times, \update{approaching} the decay rate of $1/2$ observed across all cumulants, which is consistent with the analytical predictions of \textcite{Campo_2018}.

\section{Discussions} 
Simulations of the critical dynamical behavior till a few years ago were only able in suitably 
designed laboratories. Anyway, the advent of on-demand quantum digital processing units 
made testing the universality of scaling laws (and their deviations) in complex systems possible for a much broader number of researchers. 
We perform a quantum simulation of the dynamics of a quantum phase transition using a digital superconducting quantum computer, with a circuit's depth of up to one hundred sequential CNOT gates, employing a variety of error mitigation techniques. Using error mitigation, we increase the number of reliable Trotter steps by a factor of four, at least for the first cumulant. Computing higher cumulants requires more statistics, as they are typically order of magnitude lower. Although the cost of error mitigation generally increases exponentially with the level of noise \cite{Eisert_limit_error_mitigation,complexity_QEM, schuster2024-sim-error-mitigation}, making it unsustainable in the long term, we argue that it remains a vital tool for utilizing quantum computers before achieving full fault-tolerance. Specifically, we contend that error mitigation bridges the gap towards the era of early fault-tolerance \cite{EFTQC} and facilitates the implementation of typical algorithms \cite{LT,EFTQC_practice,PRXQ_Blunt23} of this era.
\update{For physics-inspired problems, the use of error-mitigated simulations on digital devices is envisaged to fill in the gaps between universal simulation 
on classical computer clusters and special-purpose analog quantum simulators. It is worth noting that, 
the latter are restricted to realizing a model one at a time and strongly rely on a specific calibration of the device. 
Focusing on the scaling of the defects formation, the previous works conducted on analog platforms are focusing on a `quasi-annealing' 
regime. In such a case, the results are obtained over a time window in which thermal and noise fluctuations become relevant, and the 
hypothesis of closed system evolution underlying the KZM could be easily violated. 
We confirmed, in a platform-agnostic setup, that it is possible to engineer the dynamics of critical systems, 
checking the finite-size effects, paving the way to the extension for general models in which classical simulations are infeasible. 
Our paper constitutes a step into the study of complex systems, and models beyond the TFIM requiring a not 
trivial decomposition of unitary dynamics are an avenue for future simulation.}

We have empirically demonstrated in a finite-size TFQIM the deviations from the KZM during rapid 
quenches across a continuous phase transition. These deviations lead to a breakdown in the predicted power-law scaling of defect density with $\tau_Q$, resulting in a plateau preceding the $\tau_Q^{-\alpha}$ typical scaling. Additionally, we have shown how the distribution of the density of topological defects follows a binomial distribution, nearly Gaussian when the central limit theorem requirements are met \cite{Campo_23_deviation}. This model predicts that higher-order cumulants of the defect distribution scale with the quench time in a universal manner. These findings follow the pioneer line suggested by Del Campo \cite{delCampo_20, Campo_23} to test the counting statistics of topological defects and the corresponding breakdown in experimental systems.
We investigated finite-size effects using MPS by analyzing the decay rate as a function of system size, as well as recovering the full probability distribution. We observed that the decay rate converges to $1/2$ for the first three cumulants, approaching the value predicted in the thermodynamic limit and in pure coherent dynamics. 

\update{As a final remark, we would like to propel the use of digital quantum hardware as simulators of emergent phenomena. 
We have shown that even limited devices can be successfully adapted to work as versatile quantum simulators, without the requirement of having an \emph{ad hoc} laboratory. 
We do surmise that our work empirically supports the utility and advantage (for the physics community at large) 
of devices of up to thousands of physical qubits, so far from the fault-tolerance and their ambitious tasks.}
Our work goes in the direction of testing the scaling laws in quantum critical phenomena and has broad relevance in nonequilibrium statistical mechanics, numerically proving KZM, and the breakdown of adiabatic dynamics, for the statistics of the kink-density in a TFQIM. We surmise our findings are of interest to better understand the validity of quantum simulation and quantum annealing \cite{miessen2024benchmarking}, and the study of critical and thermalization phenomena in non-dissipative systems \cite{RMP_Neq, andersen2024thermalization}.

\begin{acknowledgments}
The authors thank IQM and IBM for providing access to their devices. O.K. and  M.G. are supported by CERN through the CERN Quantum Technology Initiative. 
A.M.  acknowledges that this work is partially carried out under IRA Programme, project no. FENG.02.01-IP.05-0006/23, financed by the FENG program 2021-2027, Priority FENG.02, Measure FENG.02.01., with the support of the FNP.
A.M. thanks the  CERN Theoretical Physics Department for the hospitality. 
\end{acknowledgments}

\section*{Author contributions}
A.M. designed the study. O.K. carried out the quantum simulation, while D.T. executed the tensor networks calculations. O.K. and A.M. wrote the manuscript with inputs from all authors. M.G. and A.M. supervised and coordinated the project.

\bibliography{bibliography}
\appendix

\section{Propagation of statistical uncertainties}
\label{bayesian}
\update{As described in Ref.~\cite[App. B]{PRD_neutrino}, we use a Bayesian strategy to estimate the statistical uncertainties on the mitigated results. Instead of propagating the uncertainties through the whole process, we use Bayes theorem to generate an arbitrary number of experiments, compatible with the bare data, and compute their variance. The procedure is best described considering a single qubit, whose probability of obtaining $m$ measurement of the $|1\rangle$ state out of a total of $M$ trials is given by a binominal distribution}

\begin{equation}
    P_b(m;p) = \binom{M}{m}p^m(1-p)^{(M-m)}.
\end{equation}
The probability $p$ of obtaining $|1\rangle$ is then inferred with Bayes theorem  
\begin{equation}
    P(p|m_i) = \frac{P(m_i|p)P(p)}{\int dq P(m_i|q)P(q)},
\end{equation}
which is given in closed form by the beta prior 
\begin{equation}
    P_\beta(p;\alpha,\beta) = \frac{\Gamma(\alpha+\beta)}{\Gamma(\alpha)\Gamma(\beta)}p^{(\alpha-1)}(1-p)^{(\beta-1)}.
\end{equation}
Here, $\alpha$, $\beta>0$ parameterize the beta distribution and represent the number of times the basis vector $i$ is measured. Note that we only sample from binary string appearing at least once in the data, to avoid sampling from exponentially many terms. 
After this inference phase, the following procedure can be used to determine the expected values: (i) sample a value $p_k'$ from the posterior $P(p_k'|m_i)$. (ii) sample $L$ new measurements from the likelihood $P_b(m_k';p_k')$. (iii) Compute expectations values by averaging over the generate measurements.
To generalized to multiple qubits, the Dirichlet distribution can be used, which is a closed-form prior for a multinominal distribution.

\end{document}